\documentclass[journal,twoside]{IEEEtran}
\usepackage{graphicx,amsmath,amssymb}
\usepackage[noadjust]{cite}
\usepackage{array}
\usepackage{cases}
\usepackage{url}
\usepackage{multirow}
\newtheorem{theorem}{Theorem}[section]

\newtheorem{corollary}[theorem]{Corollary}

\begin{document}
\title{High-rate self-synchronizing codes}
\author{Yuichiro~Fujiwara,~\IEEEmembership{Member,~IEEE,} and Vladimir~D.~Tonchev%
\thanks{The first author acknowledges support from JSPS Postdoctoral Fellowships for Research Abroad.
Vladimir Tonchev acknowledges the support by an NSA Grant.
The material in this paper was presented in part
at the International Symposium on Information Theory and its Applications, Honolulu, HI USA, October 2012.}%
\thanks{Y. Fujiwara is with the Division of Physics, Mathematics and Astronomy, California Institute of Technology, MC 253-37, Pasadena, CA 91125 USA
(email: yuichiro.fujiwara@caltech.edu).}
\thanks{V. D. Tonchev is with the Department of Mathematical Sciences, Michigan Technological University, Houghton, MI 49931 USA
(email: tonchev@mtu.edu).}
\thanks{Copyright \copyright\ 2012 IEEE. Personal use of this material is permitted.
However, permission to use this material for any other purposes must be obtained from the IEEE by sending a request to pubs-permissions@ieee.org.}}
\markboth{IEEE transactions on Information Theory,~Vol.~x, No.~xx,~month~year}
{Fujiwara and Tonchev: High-rate self-synchronizing codes}

%\pubid{0000--0000/00\$00.00~\copyright~2007 IEEE}

\maketitle

\begin{abstract}
Self-synchronization under the presence of additive noise can be achieved
by allocating a certain number of bits of each codeword as markers for synchronization.
Difference systems of sets are combinatorial designs which specify 
the positions of synchronization markers
in codewords in such a way that the resulting error-tolerant self-synchronizing codes may be realized as cosets of linear codes.
Ideally, difference systems of sets should sacrifice as few bits as possible
for a given code length, alphabet size, and error-tolerance capability.
However, it seems difficult to attain optimality with respect to known bounds when the noise level is relatively low.
In fact, the majority of known optimal difference systems of sets are for exceptionally noisy channels,
requiring a substantial amount of bits for synchronization.
To address this problem, we present constructions for difference systems of sets that allow for higher information rates
while sacrificing optimality to only a small extent.
Our constructions utilize optimal difference systems of sets as ingredients
and, when applied carefully, generate asymptotically optimal ones with higher information rates.
We also give direct constructions for optimal difference systems of sets with high information rates and error-tolerance
that generate binary and ternary self-synchronizing codes.
\end{abstract}


\IEEEpeerreviewmaketitle

\section{Introduction}
\IEEEPARstart{A}{}\textit{self-synchronizing code} is a block code where the 
symbol string formed by an overlapped portion of any two concatenated codewords
or a portion of any single codeword is not a valid codeword.
In the coding theory literature, self-synchronizing codes are also called \textit{comma-free codes}.
The property that no codeword appears as a substring of two adjacent codewords allows for synchronization
without any external help or separate clock signal.
Carefully designed self-synchronizing codes may be used for synchronization under the presence of additive noise as well.

Self-synchronizing codes are also of interest from mathematical viewpoints and have been investigated in both coding theory and combinatorics.
This paper focuses on a mathematical approach to the construction of self-synchronizing codes of high information rate 
by using combinatorial designs.

A {\it splice} of length $v$ between the preceding codeword $x = (x_0, x_1, \dots, x_{v-1})$ of length $v$
and the following codeword $y = (y_0, y_1, \dots, y_{v-1})$ of the same length is
a concatenated sequence $(x_{v-i}, \dots, x_{v-1}, y_0, \dots y_{v-i-1})$ composed of the last $i$ bits of $x$
and the first $v-i$ bits of $y$ for some positive integer $i \leq v-1$.

A block code $C$ of length $v$ is \textit{comma-free} with \textit{index} $\rho$
if the Hamming distance between any codeword $z$
and any splice of length $v$ between any two codewords $x, y$ is at least $\rho$.

A \textit{difference system of sets} (DSS) of \textit{index} $\rho$ over ${\textit{\textbf{Z}}}_v$
is a family of disjoint subsets $Q_i$ of ${\textit{\textbf{Z}}}_v$ such that
the multi-set 
\begin{equation}
\label{equ}
\{ a-b \pmod{v} \ \vert \ a \in Q_i, b \in Q_j, i \not=j \}
\end{equation}
 contains
every $d \in {\textit{\textbf{Z}}}_v\setminus \{0\}$ at least $\rho$ times.
The difference between two elements from different subsets of ${\textit{\textbf{Z}}}_v$ is called an \textit{outer difference}.
A DSS is {\it perfect} if the multi-set defined in (\ref{equ}) contains
every $d \in {\textit{\textbf{Z}}}_v\setminus \{0\}$ exactly $\rho$ times.
A DSS is \textit{regular} if all subsets $Q_i$ are of the same size.
A regular DSS that consists of $q$ subsets of cardinality $m$ is denoted by DSS$(v,m,q,\rho)$.

DSSs were introduced to realize self-synchronizing codes as cosets of linear codes
in order to achieve low encoding and decoding complexity \cite{L,L2}.
Regardless of which error-correcting code we use to protect the payload,
a DSS of index $\rho$ assures self-synchronization under the presence of up to $\lfloor\frac{\rho-1}{2}\rfloor$ symbol substitutions (or errors)
in the received message of length $v$.
To construct a self-synchronizing code from a DSS,
each subset $Q_i$ is used to determine which positions in a codeword are allocated for synchronization marker $i$.
For instance, the set $\{\{1,2,3,4,6,15\}, \{5,9,10,14,17,24\}\}$ forms a perfect regular DSS of index three over $\textit{\textbf{Z}}_{25}$ for a binary system.
The two sets of cardinality six specify the positions of $0$s and $1$s as synchronization markers
while the remaining thirteen positions are freely available for information transmission.
By writing a bit carrying the payload as $*$, we have 25 bit sequence
\[{*}000010{*}{*}11{*}{*}{*}10{*}1{*}{*}{*}{*}{*}{*}1.\]
Because each nonzero outer difference appears three times in the DSS,
regardless of the content of each $*$,
there are at least three discrepancies among the positions $\{1,2,3,4,6,15\}\cup\{5,9,10,14,17,24\}$
between any pair of a valid codeword of the resulting self-synchronizing code and a splice.
Hence, the receiver can tell if a given 25 bit sequence is properly block-wise synchronized
as long as the number of symbol substitution errors is at most $\lfloor\frac{3-1}{2}\rfloor = 1$.
If we use an arbitrary binary linear code of length thirteen for the free positions $*$,
the resulting self-synchronizing code is a coset of a binary linear code of length $25$.

Various construction methods for DSSs have been developed in recent years.
For the latest results and a survey of earlier research, we refer the reader to \cite{FL,ZTWY,T} and references therein.

Of particular interest are DSSs that require fewer bits for self-synchronization.
The number of bits required for synchronization is exactly the number of elements used in the DSS, that is,
$|Q_0 \cup Q_1 \cup \dots \cup Q_{q-1}|$.
All the remaining bits may freely be used for information transmission.
Hence, for example, a regular DSS$(v,m,q,\rho)$ generates a self-synchronizing code of index $\rho$
with $v-mq$ information bits.
The cardinality $|Q_0 \cup Q_1 \cup \dots \cup Q_{q-1}|$ is called the \textit{redundancy} of the DSS. 
The minimum redundancy for given $v$, $q$, and $\rho$ is denoted by $r_q(v,\rho)$.
Levenshtein \cite{L} proved the following lower bound on $r_q(v,\rho)$:
\begin{align}\label{Lbound}
r_q(v,\rho) \geq \sqrt{\frac{q\rho(v-1)}{(q-1)}}
\end{align}
with equality if and only if the DSS is perfect and regular.
A sharper bound was proved by Wang \cite{W}:
\[r_q(v,\rho) \geq \sqrt{S(\rho(v-1)+\left\lceil\frac{\rho(v-1)}{(q-1)}\right\rceil)},\]
where $S(n)$ denotes the smallest square number that is greater than or equal to $n$.

A DSS is \textit{optimal} if its redundancy is the lowest possible for given parameters.
If the redundancy of an infinite series of DSSs approaches a known lower bound as some other parameters tend to infinity,
we say that such DSSs are \textit{asymptotically optimal}.

Equally important, or perhaps more important in practical situations, is the ratio of the number of bits allocated for self-synchronization
to the number of bits available for information transmission.
In the case of a self-synchronizing code obtained from a DSS,
the \textit{redundancy rate} of a DSS over ${\textit{\textbf{Z}}}_v$ using $s$ symbols is defined as the fraction $e = \frac{s}{v}$.
For instance, the redundancy rate of a regular DSS$(v,m,q,\rho)$ is $\frac{mq}{v}$.
A DSS of redundancy rate $e$ gives a self-synchronizing code of constant code length $v$
where $ev$ bits of each codeword are used for self-synchronization.
Clearly the information rates of the corresponding self-synchronization codes
depend not on the optimality of DSSs, but on the absolute values of the redundancy rates.
For instance, a DSS that leaves only one bit for information transmission can still be judged as ``optimal" with respect to Inequality (\ref{Lbound}).
In other words, optimality means something only when the target information rate is met.

While nontrivial DSSs are of interest and may allow us to realize self-synchronizing codes as cosets of linear codes,
it seems quite difficult to construct such combinatorial objects.
A particularly difficult task is to suppress the redundancy rate to a very low level.
In fact, optimal or asymptotically optimal DSSs with redundancy rates significantly lower than a half are quite rare;
if we use a DSS of redundancy rate, say, $\frac{2}{3}$, we must sacrifice two thirds of the bits for self-synchronization,
regardless of whether it is optimal or not.
While DSSs with high redundancy rates are certainly of mathematical interest
(see, for example, \cite{CD,D,CLY,GMW,OKSS,FMY,FMT,T} for relations to other mathematical concepts),
one has to sacrifice a significant portion of bits just for synchronization.

The primary purpose of this paper is to propose a simple remedy for this information rate problem.
We give simple combinatorial methods for constructing DSSs with lower redundancy rates from ones with higher redundancy rates,
allowing for self-synchronizing codes with improved information rates.
Our methods give asymptotically optimal DSSs with low redundancy rates when applied to carefully chosen ingredients.
We also present direct constructions for optimal DSSs that are suited for use as ingredients of our redundancy reduction methods.
Discussed at the end of this paper are some open problems and directions of research on DSSs which the authors believe are important.
Table \ref{table} in Appendix summarizes the parameters of perfect regular DSSs given in this paper
for the case when the corresponding self-synchronizing codes are binary or ternary
and of synchronization redundancy rate strictly smaller than $\frac{1}{2}$.
All known series of perfect regular DSSs of redundancy rate less than a half are also listed for convenience.

\section{Product constructions}
In this section, we give combinatorial constructions that generate difference systems of sets with lower redundancy rates from those with higher redundancy rates.
For the sake of simplicity, for the most part we use perfect regular DSSs as ingredients to derive new DSSs.
The constructions can generate various infinite classes of difference systems of sets including asymptotically optimal ones.
The same technique is applicable to any DSS that may or may not be perfect or regular in a straightforward manner.

To make it easier to see the mechanism of the redundancy reduction process,
we first give a simpler construction for DSSs
\footnote{Essentially the same construction appeared without proof
in the context of combinatorics of outer differences in a workshop abstract by the first author and Fuji-Hara \cite{FF} (see also \cite{WW}).
Here we give a complete proof and explain how this technique improves information rates in the context of self-synchronizing codes.}
and demonstrate how to use it to reduce the redundancy rates of known DSSs.
A generalized version of the construction is then presented.

Let $\mathcal{P}$ be a family of subsets $B_0, B_1, \dots, B_{q-1}$ of size $m$ over ${\textit{\textbf{Z}}}_v$.
The family ${\cal P}$ is said to form a {\it difference family} over ${\textit{\textbf{Z}}}_v$
and is denoted by DF$(v,m,\lambda)$ if every nonzero element of ${\textit{\textbf{Z}}}_v$ appears exactly $\lambda$ times in
the multi-set $\{a - b \ \vert \ a, b \in B_i, 0 \leq i \leq q-1\}$.
A difference family can be defined the same way when $\mathcal{P}$ may contain subsets of different cardinalities.
If the subset sizes are not uniform, we specify possible sizes by the set $K = \{\vert B_i \vert\ \vert\ 0 \leq i \leq q-1\}$ and write DF$(v,K,\lambda)$.
When $\mathcal{P}$ consists of a single set of size $m$, it is a cyclic 
\textit{difference set} and written as DS$(v,m,\lambda)$.
As opposed to outer differences, the difference between a pair of elements of the same set is called an \textit{inner difference}.
Roughly speaking, a DF is an inner version of a perfect DSS
in the sense that the number of occurrences of each inner difference in a DF is uniform across all the nonzero elements
while in a perfect DSS nonzero outer differences occur uniformly.

\begin{theorem}\label{construction}
Let $v$ and $v'$ be relatively prime positive integers.
If there exist a perfect regular \textup{DSS}$(v,m,q,\rho)$ forming a \textup{DF}$(v,m,\lambda)$ and a perfect regular \textup{DSS}$(v',m',q',\rho')$ forming
a \textup{DF}$(v',m',\lambda')$, then there exists a regular \textup{DSS}$(vv',mm',qq',\min(\rho\rho'+\rho\lambda'+\rho'\lambda, \rho m'q', \rho'mq))$.
\end{theorem}
\begin{IEEEproof}
Let $\mathcal{A} = \{Q_0, Q_1, \dots, Q_{q-1}\}$ and $\mathcal{B} = \{Q'_0, Q'_1, \dots, Q'_{q'-1} \}$ be
a perfect regular DSS$(v,m,q,\rho)$ forming a \textup{DF}$(v,m,\lambda)$ and
a perfect regular DSS$(v',m',q',\rho')$ forming a DF$(v',m',\lambda')$, respectively.
Take family
\[\mathcal{C} = \{ Q_i \times Q'_j \ \vert \ 0 \leq i \leq q-1, 0 \leq j \leq q'-1 \}\]
of all the direct products between elements of $\mathcal{A}$ and those of $\mathcal{B}$.
Since $v$ and $v'$ are relatively prime, $\mathcal{C}$ can be seen as a family of $qq'$ disjoint sets of size $mm'$ over ${\textit{\textbf{Z}}}_{vv'}$.
It suffices to prove that each outer difference appears either $\rho\rho'+\rho\lambda'+\rho'\lambda$, $\rho m'q'$ or $\rho'mq$ times.

Write the elements of the cyclic group of order $vv'$ as $(a,b)$, where $a \in {\textit{\textbf{Z}}}_{v}$ and $b \in {\textit{\textbf{Z}}}_{v'}$.
Since $\mathcal{A}$ is a family of disjoint sets, an outer difference of the form $(0,b)$
only occurs between $Q_i \times Q'_j$ and $Q_i \times Q'_k$ for some $j$ and $k$.
Assume that $Q'_j$ and $Q'_k$ give $b$ as an outer difference exactly $x_{j,k}$ times.
For every $i$, $0 \leq i \leq q-1$, and fixed $j$ and $k$,
there are $m \cdot x_{j,k}$ instances of outer difference $(0,b)$ between $Q_i \times Q'_j$ and $Q_i \times Q'_k$.
Since $\mathcal{B}$ is a DSS of index $\rho'$, taking all possible pairs $Q'_j$ and $Q'_k$ gives the outer difference $b$ $\rho'$ times.
Hence, we have

\[\sum_{i}\sum_{j,k} m \cdot x_{j,k} = \rho'mq.\]

Hence, we have each outer difference of the form $(0,b)$ exactly $\rho'mq$ times in $\mathcal{C}$.
By the same token, each outer difference of the form $(a,0)$ occurs exactly $\rho m'q'$ times.

Consider an outer difference of the form $(a,b)$ with $a, b \not= 0$.
We first consider the outer differences between the elements in $Q_i \times Q'_j$ and those in $Q_i \times Q'_k$ with $j \not= k$.
Since $\mathcal{A}$ is a DF$(v,m,\lambda)$, inner difference $a$ occurs exactly $\lambda$ times in $\mathcal{A}$.
For each occurrence, taking all possible $j$ and $k$ gives $\rho'$ instances of outer difference $(a,b)$.
Hence, for fixed $a$ and $b$, we have $(a,b)$ as an outer difference exactly $\rho'\lambda$ times
between $Q_i \times Q'_j$ and $Q_i \times Q'_k$ with $j \not= k$.
Similarly, we have $\rho\lambda'$ $(a,b)$s between $Q_i \times Q'_j$ and $Q_k \times Q'_j$ with $i \not= k$.
Consider outer differences between $Q_i \times Q'_j$ and $Q_k \times Q'_l$ with $i \not=k$ and $j \not= l$.
Since $\mathcal{A}$ and $\mathcal{B}$ are DSSs of indices $\rho$ and $\rho'$ respectively, it is straightforward to see that
by taking all possible $i$, $j$, $k$, $l$, we get $(a,b)$ exactly $\rho\rho'$ times
between $Q_i \times Q'_j$ and $Q_k \times Q'_l$ with $i \not=k$ and $j \not= l$.
Hence we have each outer difference of the form $(a,b)$ with $a, b \not=0$ exactly $\rho\rho'+\rho\lambda'+\rho'\lambda$ times.
The proof is complete.
\end{IEEEproof}

Note that the redundancy rate of the resulting DSS is the product between those of the DSSs used as ingredients.
Because the redundancy rate of any DSS is less than or equal to $1$,
the resulting DSS always has a lower or equal redundancy rate when compared to the ingredients,
which means that the corresponding self-synchronizing code can take advantage of more information bits.
It is also worth noting that the same technique can be applied to DSSs that do not form difference families,
albeit with a more complicated analysis of the number of occurrences of each inner and outer difference.

The direct product technique described above can give infinitely many series of asymptotically optimal regular difference systems of sets.
The following is an example of an infinite class of such DSSs obtained from the ones of Paley type (see \cite{T2}):

\begin{corollary}\label{coroPaley}
Let $v$ and $v'$ be two distinct primes congruent to $3$ modulo $4$ and write $v=2mq+1$ and $v'=2m'q'+1$
for some positive integers $m$, $m'$, $q$, and $q'$, respectively.
Then there exists an asymptotically optimal class of regular \textup{DSS}s of parameters
$(vv'\hspace{-0.24mm},mm'\hspace{-0.24mm},qq'\hspace{-0.24mm},\hspace{-0.37mm}\frac{m(m'-1)(q-1)+(m-1)m'(q'-1)+mm'(q-1)(q'-1)}{4}\hspace{-0.1mm})$.
\end{corollary}

\begin{IEEEproof}
Let $v$ and $v'$ be two distinct primes congruent to $3$ modulo $4$ as stated in the statement.
Then there exist prefect regular DSSs of parameters $(v,m,q, \frac{v-2m-1}{4})$ and $(v',m',q', \frac{v'-2m'-1}{4})$
which form DFs of indices $\lambda = \frac{m-1}{2}$ and $\lambda' = \frac{m'-1}{2}$, respectively \cite{T2}.
Let $\rho = \frac{v-2m-1}{4}$ and $\rho' = \frac{v'-2m'-1}{4}$.
A simple calculation of the comma index shows that
$\min(\rho\rho'+\rho\lambda'+\rho'\lambda, \rho m'q', \rho'mq) = \rho\rho'+\rho\lambda'+\rho'\lambda\\
= \frac{m(m'-1)(q-1)+(m-1)m'(q'-1)+mm'(q-1)(q'-1)}{4}$.
Applying Theorem \ref{construction} gives a regular DSS of the desired parameters.
By using Inequality (\ref{Lbound}), it is straightforward to show that
\[\lim_{m,m' \rightarrow \infty} \frac{r_{qq'(vv', \rho\rho'+\rho\lambda'+\rho'\lambda)}}{mm'qq'} = 1.\]
The proof is complete.
\end{IEEEproof}

The asymptotically optimal DSSs allow for greatly improved information rates compared to the ingredient systems.
In fact, the redundancy rate of a DSS obtained from
Corollary \ref{coroPaley} is only $\frac{(v-1)(v'-1)}{4vv'} \approx \frac{1}{4}$ while that of the Paley type DSSs is $\frac{v-1}{2v} \approx \frac{1}{2}$.

The redundancy rates of the resulting DSSs in Theorem \ref{construction} depend on ingredient systems.
Hence, direct constructions of DSSs having low redundancy rates are important in constructing a DSS with very low redundancy rates.
Among many results in the literature,
perfect regular DSSs with remarkably low redundancy rates were given in \cite{FMT}
by partitioning the points of hyperplanes of projective spaces:

\begin{theorem}[\cite{FMT}]\label{hyperplaneDSS}
There exists a partition of the points of a hyperplane of the projective space \textup{PG}$(2s,p)$ into a perfect regular
\textup{DSS}$(\frac{p^{2s+1}-1}{p-1},p+1,\frac{p^{2s}-1}{p^2-1},\frac{p^{2s-1}-p}{p-1})$ forming a \textup{DF}$(\frac{p^{2s+1}-1}{p-1},p+1,1)$ for
$p=2, 3, 5, 8, 9$ and $s=2$, and $p = 2, 3$ and $s = 3$, and $p = 2$ and $s = 4, 5$.
\end{theorem}

For example, the redundancy rate of their optimal DSS from PG$(4,9)$ is $\frac{9^4-1}{9^5-1} \approx \frac{1}{9}$.
Theorem \ref{construction} can lower this rate even further.
For instance, we can reduce the redundancy rate to approximately $\frac{1}{72}$
by applying the direct product technique with the DSS from PG$(4,8)$.

\begin{corollary}\label{corohyperplane}
Let $p$ and $s$ be positive integers greater than one satisfying $p=2$ and $s \leq 5$, $p=3$ and $s \leq 3$ or $p=5, 8, 9$ and $s=2$.
Take one more pair $p'$ and $s'$ of integers satisfying the same condition.
If $\gcd(\frac{p^{2s+1}-1}{p-1},\frac{p'^{2s'+1}-1}{p'-1}) = 1$, then
there exists a regular \textup{DSS}$(\frac{(p^{2s+1}-1)(p'^{2s'+1}-1)}{(p-1)(p'-1)},(p+1)(p'+1),\frac{(p^{2s}-1)(p'^{2s'}-1)}{(p^2-1)(p'^2-1)},
\frac{(p^{2s-1}-1)(p'^{2s'-1}-1)}{(p-1)(p'-1)}-1)$.
\end{corollary}
\begin{IEEEproof}
Apply Theorem \ref{construction} to Theorem \ref{hyperplaneDSS}.
The assertion follows from a simple calculation.
\end{IEEEproof}

We now generalize the construction technique used in Theorem \ref{construction}.
The previous construction requires that the lengths of the pair of self-synchronizing codes
corresponding to the DSSs used as ingredients be relatively prime.
We relax this condition by using a combinatorial technique similar 
to the one found in \cite{DFFJM}.
Unlike the simpler construction, the generalized version does not simply take 
the direct product between two sets from a pair of DSSs.
To avoid unduly involved technical arguments and succinctly present 
the combinatorics behind the key idea,
we restrict one ingredient to a DSS of redundancy rate one.
Such DSSs are equivalent to frequency hopping patterns for spread-spectrum multiple access communications.
More formally, a \textit{frequency hopping sequence} of \textit{period} $v$ over a set $F$ of cardinality $q$
is a $v$-dimensional vector $X = (x_0, x_1, \dots, x_{v-1})$ with $x_i \in F$ for $0 \leq i \leq v-1$, where $\vert F \vert = q$.
By taking the support of each element of $F$,
we obtain $q$ disjoint subsets partitioning the set $\{0,1,\dots,v-1\}$, which can be seen as a DSS of certain index.

One objective of the study of frequency hopping sequences is to minimize the number of occurrences of each inner difference for given $v$ and $q$,
or equivalently, to minimize the off-peak Hamming autocorrelations for given $v$ and $q$ (see \cite{FMM}).
It is straightforward to see that the sum of the number of occurrences of inner difference $i$ and that of outer difference $i$ is $v$.
Hence, a DSS of index $\rho$ and redundancy rate one is equivalent to
a frequency hopping sequence in which each inner difference appears at most $v - \rho$ times.
In what follows, we write a DSS of index $\rho$ and redundancy rate one on $q$ sets over $\textit{\textbf{Z}}_v$ as FHS$(v,v-\rho; q)$.

\begin{theorem}\label{construction2}
If there exist an \textup{FHS}$(v,v-\rho; q)$ and
a perfect \textup{DSS} of index $\rho'$ and redundancy rate $e'$ over $\textit{\textbf{Z}}_v'$ forming a \textup{DF}$(v',K,\lambda')$,
then there exists a \textup{DSS} of index $\min(\rho e'v', v(\lambda'+\rho'))$ and redundancy rate $e'$ on $qv'e'$ sets over $\textit{\textbf{Z}}_{vv'}$.
\end{theorem}
\begin{IEEEproof}
Let $\mathcal{A} = \{Q_0, Q_1, \dots, Q_{q-1}\}$ and $\mathcal{B} = \{Q'_0, Q'_1, \dots, Q'_{q'-1} \}$ be
an FHS$(v,v-\rho; q)$ and a DSS with the parameters given in the statement respectively.
We write the elements of the rings ${\textit{\textbf{Z}}}_v$ and ${\textit{\textbf{Z}}}_v'$ by
$\{0,1,\dots,v-1\}$ and $\{0,1,\dots,v'-1\}$ respectively.
We construct subsets of ${\textit{\textbf{Z}}}_{vv'}$ by embedding the elements of the two rings.
For every $Q_i$ and $x \in \bigcup_j Q'_j$, define the set $S_{i,x} = \{v'a+x \ \vert \ a \in Q_i\}$ over ${\textit{\textbf{Z}}}_{vv'}$.
Let
\[\mathcal{S} = \left\{S_{i,x} \ \middle\vert \ 0 \leq i \leq q-1, x \in \bigcup_j Q'_j\right\}.\]
$\mathcal{S}$ is a family of disjoint $qv'e'$ subsets of ${\textit{\textbf{Z}}}_{vv'}$.
We have $\vert\bigcup S_{i,x}\vert = e'vv'$.
It suffices to prove that each outer difference in $\mathcal{S}$ appears at least $\min(\rho e'v', v(\lambda'+\rho'))$ times.

An outer difference that is divisible by $v'$ appears at least $\rho$ times between $S_{i,x}$ and $S_{j,x}$.
Because there are $v'e'$ choices for $x$, the number of occurrences of an outer difference of this kind is at least $\rho v'e'$.
An outer difference that is not divisible by $v'$ appears exactly $\lambda' v$ times between $S_{i,x}$ and $S_{j,y}$ for $x, y \in Q'_k$,
and $\rho' v$ times between $S_{i,x}$ and $S_{j,y}$ for $x \in Q'_k, y \in Q'_l$ with $k \not= l$.
Hence, the total number of occurrences is $v(\lambda'+\rho')$.
The proof is complete.
\end{IEEEproof}

Frequency hopping sequences have extensively been studied from various viewpoints.
Constructions for frequency hopping sequences with optimal Hamming autocorrelations can be found in
\cite{LG,Kumar,KL,US,FMM,CC2,GFM,DMY,HY,CHY,CY,CY2,ZCTY}.
Known constructions for sets of frequency hopping sequences may be used for Theorem \ref{construction2} as well
because each set contains frequency hopping sequences with good Hamming autocorrelations (see \cite{ZTPP,YTPP} for recent results).
Equivalent or closely related mathematical objects have also been investigated under the names of
constant composition codes \cite{CCD,DY,LFVC} (see also \cite{LH,ZG} for more details and the latest results),
partition difference families \cite{YST,WW2,LF}, external difference families \cite{CD,HW},
and zero-difference balanced functions \cite{Ding,ZTWY}.

In the remainder of this section, we briefly look into what kind of DSS can be obtained through the technique used in Theorem \ref{construction2}.

As in Theorem \ref{construction}, the redundancy rate of the resulting difference system of sets generated
by the technique given in the proof of Theorem \ref{construction2} is the product of the redundancy rates of the two ingredients.
Hence, a DSS of extremely high redundancy rate will not lead to a significantly improved information rate.
In this sense, it is important to utilize at least one DSS of low redundancy rate as an ingredient.
Nonetheless, a frequency hopping sequence, which is a DSS that uses up all bits, can still be used to obtain DSSs
of very good or even optimal redundancy with respect to Inequality (\ref{Lbound}).
In fact, Theorem $4$ in \cite{WW} can be seen as a corollary of Theorem \ref{construction2}:
\begin{corollary}\label{coroDS}
If there exist an \textup{FHS}$(v,v-\rho; q)$ and
a \textup{DS}$(v',m',\lambda')$,
then there exists a \textup{DSS} of  index $\min(\rho m', v\lambda')$ and redundancy rate $\frac{m'}{v'}$ over $\textit{\textbf{Z}}_{vv'}$
based on $qm'$ sets.
\end{corollary}
\begin{IEEEproof}
A DS$(v',m',\lambda')$ is also a perfect regular DSS$(v',m',1,0)$ forming a DF$(v',m',\lambda')$.
Applying Theorem \ref{construction2} proves the assertion.
\end{IEEEproof}

Difference sets are important combinatorial objects and have been a topic of extensive research \cite{HandbookCD}.
To see how good the DSSs of Corollary \ref{coroDS} are in terms of optimality,
take, for example, the projective plane over the finite field of order $k$ as a DS$(k(k-1)+1,k,1)$.
If we fix the FHS$(v,v-\rho; q)$ used as an ingredient, the index of the resulting DSS is $\min(\rho k, v) = v$ for large $k$.
The ratio between the redundancy of the resulting DSS and the right-hand side of Inequality (\ref{Lbound}) approaches $1$ as $k$ tends to infinity.
Hence, we obtain an infinite series of asymptotically optimal DSSs.
The construction process of the optimal DSS of redundancy rate $\frac{3}{7}$ over $\textit{\textbf{Z}}_{49}$ given in Example $5$ of \cite{WW}
can be seen as an application of the Fano plane to Corollary \ref{coroDS}.

In general, the product techniques only slightly, if at all, degrade optimality
if ingredients are chosen so that every outer difference appears almost uniformly in the resulting DSS.
One possible drawback of the product constructions is that the code length is inherently longer than those of the codes used as ingredients.
This implies that a high-rate self-synchronizing code of very short length is difficult to obtain by our approach.
Another restriction on the available lengths is that they must be composite numbers,
which can be a problem if one wishes a code of prime length.
The increased alphabet size may also be of concern if one would like to employ self-synchronizing codes in a $q$-ary system with very small $q$.
We deal with these problems in the following section by giving direct constructions for binary and ternary DSSs over prime fields.

\section{Cyclotomic constructions}

To take advantage of the techniques presented in the previous section,
we need difference systems of sets with good parameters to start with.
In the context of improving information rates,
generally speaking, DSSs with low redundancy rates are desirable as ingredients.
Perfect regular DSSs are particularly suited for this task
because they make it easier to calculate the parameters of the resulting DSSs
while ensuring low redundancy due to their optimality guaranteed by the fact that
the equality in (\ref{Lbound}) holds if and only if a DSS is simultaneously perfect and regular.

In this section we give perfect regular DSSs of redundancy rate less than $\frac{1}{2}$.
To this end, we revisit a known direct construction for DSSs based on cyclotomy \cite{MT}.
Although it is known that DSSs of various types and paremeters can be constructed in a similar manner \cite{PCMutoh}
\footnote{In fact, the constructions given in this section may be regarded as special cases of the results reported in an unpublished manuscript \cite{Mutoh}.},
to keep clarity and simplicity of our approach,
we focus on the kind of DSS that is particularly suited for our purpose
and do not deal with DSSs that would be too cumbersome to apply to the product constructions.
For convenience, the parameters of perfect regular DSSs constructed in this section and known such systems are listed in Appendix.

Let $p = fm+1$ be an odd prime for some positive integers $f$ and $m$.
The $f$th \textit{cyclotomic classes} in $\mathbb{F}_p$ are defined as
$C_i^f = \{\alpha^{i+tf} \ \vert \ 0 \leq t \leq m-1\}$, where $\alpha$ is a primitive element of $\mathbb{F}_p$
and $0 \leq i \leq f-1$.
The \textit{cyclotomic numbers} of \textit{order} $f$ are $(i,j)_f = \left\vert (C_i^f+1) \cup C_j^f\right\vert$.
We use the following theorem:

\begin{theorem}[\cite{MT}]\label{cyclotomic}
Let $p =fmq+1$ be an odd prime, where $f$, $m$, and $q$ are positive integers.
The family $\{C_{fi}^{fq} \ \vert \ 0 \leq i \leq q-1\}$ of cyclotomic classes is a regular \textup{DSS}$(p,m,q,\rho)$, where
\[\rho = \min\left(\sum_{j=0}^{q-1}\sum_{a=1}^{q-1}(i+jf, af)_{fq} \ \middle\vert \ 0 \leq i \leq f-1\right).\]
In particular, if
\[\sum_{j=0}^{q-1}\sum_{a=1}^{q-1}(i+jf, af)_{fq} = \frac{m(q-1)}{f}\]
for every $i$, then the regular \textup{DSS} is of index $\frac{m(q-1)}{f}$, perfect, and hence optimal.
\end{theorem}

Note that Theorem \ref{cyclotomic} was originally stated in a slightly different way.
A simple calculation of cyclotomic numbers gives the above form.

A few sporadic examples of perfect regular DSSs were found through Theorem \ref{cyclotomic} \cite{MT}.
Our key observation here is that in some cases it is readily checked whether 
the condition
\[ \sum_{j=0}^{q-1}\sum_{a=1}^{q-1}(i+jf, af)_{fq} = \frac{m(q-1)}{f} \]
holds for every $i$,
so that the cyclotomic construction can give a series of perfect regular DSSs with low redundancy rates.

\begin{theorem}\label{half1}
For every $n$ such that $16n^2+1$ is an odd prime,
there exists a perfect regular \textup{DSS}$(16n^2+1,4n^2,2,2n^2)$.
\end{theorem}
\begin{IEEEproof}
Assume that $16n^2+1$ is an odd prime.
Take $C_0^4$ and $C_2^4$.
Because $\frac{(16n^2+1) -1}{4} = 4n^2$ is even,
by the classic result on cyclotomic numbers for when the order is a small 
divisor of $p-1$, $p$ prime \cite{Dickson},
we have
\begin{align*}
(0, 2)_{4} + (2, 2)_{4} &= 2(0,2)_{4}\\
&= \frac{16n^2+1-3+2}{8}\\
&= 2n^2
\end{align*}
and
\begin{align*}
(1, 2)_{4} + (3, 2)_{4} &= 2(1,2)_{4}\\
&= \frac{16n^2+1+1-2}{8}\\
&= 2n^2.
\end{align*}
Thus, we have
\begin{align*}
(0, 2)_{4} + (2, 2)_{4} &= (1, 2)_{4} + (3, 2)_{4}\\
&= 2n^2\\
&= \frac{4n^2(2-1)}{2}.
\end{align*}
Applying Theorem \ref{cyclotomic} completes the proof.
\end{IEEEproof}

Because the DSSs in Theorem \ref{half1} are both perfect and regular, they are optimal
\footnote{During revision we found that the DSSs of Paley-type (see \cite{T2}) were rediscovered in \cite{HW}
as disjoint difference families that simultaneously form external difference families,
and that the parameters realized in Theorem \ref{half1} were also independently discovered in a similar fashion by the same authors.}.
The redundancy rate is $\frac{8n^2}{16n^2+1} \approx \frac{1}{2}$.

This technique works for primes of other similar forms as well.
Here we give two more example series of perfect regular DSSs,
one of which gives redundancy rate about $\frac{1}{2}$ and the other about $\frac{1}{3}$.
The former generates optimal ternary DSSs, and the latter binary.

\begin{theorem}\label{half2}
For every $n$ such that $12n^2+1$ is an odd prime,
there exists a perfect regular \textup{DSS}$(12n^2+1,2n^2,3,2n^2)$.
\end{theorem}
\begin{IEEEproof}
Take positive integer $n$ such that $12n^2+1$ is an odd prime.
Take $C_0^6$, $C_2^6$, and $C_4^6$.
By the same argument as in the proof of Theorem \ref{half1}, we have
\begin{align*}
\sum_{j=0}^{2}\sum_{a=1}^{2}(0+2j, 2a)_{6} &= 2\left((0,2)_6 + (0,4)_6 + (2,4)_6\right)\\
&= \frac{12n^2+1-3+2}{6}\\
&= 2n^2
\end{align*}
and
\begin{align*}
\sum_{j=0}^{2}\sum_{a=1}^{2}(1+2j, 2a)_{6} &= 2\left((1,2)_6 + (1,3)_6 + (1,4)_6\right)\\
&= \frac{12n^2+1+1-2}{6}\\
&= 2n^2.
\end{align*}
Hence, by Theorem \ref{cyclotomic} the cyclotomic classes form a perfect regular DSS as desired.
\end{IEEEproof}

\begin{theorem}\label{third}
For every $n$ such that $108n^2+1$ is an odd prime,
there exists a perfect regular \textup{DSS}$(108n^2+1,18n^2,2,6n^2)$.
\end{theorem}
\begin{IEEEproof}
Let $p = 108n^2+1$ be an odd prime.
Take $C_0^6$ and $C_3^6$.
Because $2$ is a cubic residue of $p$, as in the proofs of the previous two theorems, we have
\[(i, 3)_{6} + (i+3, 3)_{6} = \frac{18n^2(2-1)}{3}\]
for $i = 0, 1$.
\end{IEEEproof}

Whether Theorems \ref{half1}, \ref{half2}, and \ref{third} are infinite series of optimal DSSs
depends on whether there exist infinitely many primes of the form $an^2+b$ for given $a$ and $b$.
The simplest case when $a = b = 1$ is already a notoriously difficult problem, 
known as Landau's problem, which has been open for a hundred years.
Nonetheless, these appear to be good sources of perfect regular DSSs with low redundancy rates on only two or three sets,
which are quite rare in the literature.
The parameters of the perfect regular DSSs given in Theorems \ref{half1}, \ref{half2}, and \ref{third} for $n \leq 10$ are listed in Table \ref{numerical}.
\begin{table}
\renewcommand{\arraystretch}{1.3}
\caption{Perfect regular DSS$(v,m,q,\rho)$s from cyclotomic constructions for $n \leq 10$}
\label{numerical}
\centering
\begin{tabular}{ccccccc}
\hline\hline
$n$ & $v$ & $m$ & $q$ & $\rho$ & $\frac{mq}{v}$ & \bfseries Reference\\
\hline
$1$ & $17$ & $4$ & $2$ & $2$ & $\frac{8}{17}$ & Theorem \ref{half1}\\
$4$ & $257$ & $64$ & $2$ & $32$ & $\frac{128}{257}$ & Theorem \ref{half1}\\
$5$ & $401$ & $100$ & $2$ & $50$ & $\frac{200}{401}$ & Theorem \ref{half1}\\
$6$ & $577$ & $144$ & $2$ & $72$ & $\frac{288}{577}$ & Theorem \ref{half1}\\
$9$ & $1297$ & $324$ & $2$ & $162$ & $\frac{648}{1297}$ & Theorem \ref{half1}\\
$10$ & $1601$ & $400$ & $2$ & $200$ & $\frac{800}{1601}$ & Theorem \ref{half1}\\
\hline
$1$ & $13$ & $2$ & $3$ & $2$ & $\frac{6}{13}$ & Theorem \ref{half2}\\
$3$ & $109$ & $18$ & $3$ & $18$ & $\frac{54}{109}$ & Theorem \ref{half2}\\
$4$ & $193$ & $32$ & $3$ & $32$ & $\frac{96}{193}$ & Theorem \ref{half2}\\
$6$ & $433$ & $72$ & $3$ & $72$ & $\frac{216}{433}$ & Theorem \ref{half2}\\
$8$ & $769$ & $128$ & $3$ & $128$ & $\frac{384}{769}$ & Theorem \ref{half2}\\
$10$ & $1201$ & $200$ & $3$ & $200$ & $\frac{600}{1201}$ & Theorem \ref{half2}\\
\hline
$1$ & $109$ & $18$ & $2$ & $6$ & $\frac{36}{109}$ & Theorem \ref{third}\\
$2$ & $433$ & $72$ & $2$ & $24$ & $\frac{144}{433}$ & Theorem \ref{third}\\
$6$ & $3889$ & $648$ & $2$ & $216$ & $\frac{1296}{3889}$ & Theorem \ref{third}\\
 \hline
 \hline
\end{tabular}
\end{table}

The DSSs given in this section are optimal and have very small $q$ and relatively low redundancy rates.
Theorem \ref{cyclotomic} can give many more perfect and almost perfect DSSs in a similar way by computing cyclotomic numbers.
If one wishes to further reduce redundancy rates by the product constructions,
the comma-free indices of the resulting DSSs can be calculated by the indices of the ingredients
and the number $\min_i\left(\sum_{j=0}^{q-1}(i+jf, 0)_{fq}\right)$ of appearances of the least frequent inner difference in each ingredient (see \cite{MT}).
Hence, while it seems impossible to give a simple and general 
formula for the exact values of the parameters of the resulting DSSs obtained 
in this manner,
calculating them for each individual case is relatively easy.

\section{Conclusion}
We have developed simple combinatorial methods for reducing the redundancy rates of difference systems of sets
while sacrificing optimality to only a small extent.
In fact, our product constructions give asymptotically optimal DSSs when applied to carefully chosen optimal DSSs.
This provides a simple remedy for the problem that even optimal DSSs may end up using a significant portion of bits
which otherwise could be used for information transmission.
Our methods hence improve the information rate of communications
while allowing for a systematic construction for self-synchronizing codes of low redundancy.
To take full advantage of and complement our methods, we also constructed perfect regular DSSs with low redundancy rates
directly through cyclotomy.
While we focused on the kind of DSS that can not be obtained through the product constructions and is useful for our approach to improving information rates,
the cyclotomic construction can give various series of regular DSSs with excellent redundancy that are of interest on their own.
A further look into this type of construction would be interesting.

As far as the authors are aware, the result presented here is the first mathematical approach that draws attention
to systematically lowering the redundancy rates of DSSs and improving the information rates of
the corresponding self-synchronizing codes.
Because the absolute values of redundancy rates are as important as optimality,
we believe that further investigations into redundancy rates are needed from both theoretical and practical viewpoints.

\subsection*{Open problems}
Among many open problems in the study of difference systems of sets,
a particularly important one would be to find explicit constructions for optimal or almost optimal DSSs with prescribed error tolerance capacities.
Regardless of optimality, DSSs with error tolerance higher or lower than the noise level of the channel would be less desirable;
they either eat up too many bits in a codeword if error tolerance is too high or do not offer secure synchronization if it is too low.
Because known optimal DSSs often have quite large $\rho$,
constructions for DSSs of small index would be of interest.

An equally important parameter is the number of sets in a DSS.
A DSS on $q$ sets gives a self-synchronizing code for a $q$-ary system if used in a straightforward manner.
A useful observation is that a DSS on $q$ sets may be used to construct a $q'$-ary self-synchronizing code for any $q' \geq q$
because there is no need to use each and every available symbol as a synchronization marker.
In fact, simply using an arbitrary $q'$-ary block code in place of the $q$-ary code for the payload gives a $q'$-ary self-synchronizing code.
It would be natural to look into the case when $q$ is a small prime or prime power,
as such DSSs would be more versatile and suited for when the information bits are protected by linear codes.

Finally, arguably the most important open problem from a more theoretical viewpoint
is to determine the asymptotic behavior of minimum redundancy $r_q(v,\rho)$.
Levenshtein \cite{L} proved that
\[r_2(v,1) = \left\lceil \sqrt{2(v-1)} \right\rceil\]
and that
\[r_2(v,2) = \left\lceil 2\sqrt{v-1} \right\rceil.\]
However, we need index to be at least three to have error tolerance.
Unfortunately, little is known about asymptotic behavior of $r_q(v,\rho)$ for $\rho \geq 3$
(see Levenshtein \cite{L2} and references therein for the background of the study in this direction).
We believe that the most important problem is to determine $r_q(v,\rho)$ for small $q \geq 2$ and $\rho \geq 3$
or its asymptotic behavior,
preferably through giving explicit constructions.

\appendix[Table of optimal difference systems of sets]
Here we list the parameters of new and known series of difference systems of sets with redundancy rates strictly less than $\frac{1}{2}$
which are perfect, regular, and error-tolerant, that is, of index at least three.
The DSSs in Table \ref{table} are sorted in order of base $q$ because of the important fact that
the corresponding self-synchronizing codes can be realized as $q'$-ary codes for any $q' \geq q$.
For completeness, two series from unpublished material \cite{Mutoh} are included at the end.
For the sake of readability, however, the table does not include asymptotically optimal series, sporadic examples,
or optimal DSSs that are either imperfect or irregular.
While we also excluded perfect regular DSSs with redundancy rates exactly $\frac{1}{2}$ from our table, such systems can be found in \cite{LF}.
Explicit examples of optimal binary, ternary and quaternary DSSs over $\textit{\textbf{Z}}_v$ with $v \leq 30$
discovered by a computer search can be found on the second author's website \cite{web}.

\begin{table*}
\renewcommand{\arraystretch}{1.3}
\caption{Series of perfect regular DSSs of redundancy rate less than a half}
\label{table}
\centering
\begin{tabular}{ccccccc}
\hline\hline
\bfseries Length $v$ & \bfseries Set size $m$ & \bfseries Base $q$ & \bfseries Index $\rho$ & \bfseries Redundancy rate & \bfseries Constraint & \bfseries Reference\\
\hline
$108n^2+1$ & $18n^2$ & $2$ & $6n^2$ & $\frac{36n^2}{108n^2+1}$ & $108n^2+1$ prime& Theorem \ref{third}\\
$16n^2+1$ & $4n^2$ & $2$  & $2n^2$ & $\frac{8n^2}{16n^2+1}$ & $16n^2+1$ prime& Theorem \ref{half1}\rlap{\textsuperscript{a}}\\
$12n^2+1$ & $2n^2$ & $3$  & $2n^2$ & $\frac{6n^2}{12n^2+1}$ & $12n^2+1$ prime& Theorem \ref{half2}\\
\hline
\multirow{3}{*}{$12n+7$} & \multirow{3}{*}{$2n+1$} & \multirow{3}{*}{$3$} & \multirow{3}{*}{$2n+1$} & \multirow{3}{*}{$\frac{6n+3}{12n+7}$} &
$12n+7 = x^2+3y^2 \geq 19$ prime, & \multirow{3}{*}{Theorem 3.4 \cite{HW}}\\
 & &  & & & $48n+28 = a^2+3b^2 = c^2+27d^2$, & \\
 & &  & & & $2x = a, 2y = b$. & \\
$4n+3$ & $m$& $\frac{2n+1}{m}$ & $\frac{2n+1-m}{2}$ & $\frac{2n+1}{4n+3}$ & $4n+3$ prime, $m \mid 2n+1$ & Theorem 2.4 \cite{T2}\rlap{\textsuperscript{a}}\\
$4n+1$ & $2$ & $n$ & $n-1$ & $\frac{2n}{4n+1}$ & $4n+1$ prime & Lemma 19 \cite{CD}\\
$8n+1$ & $4$ & $n$ & $2n-2$ & $\frac{4n}{8n+1}$ & $8n+1$ prime & Proposition 21 \cite{CD}\\
\multirow{2}{*}{$12n+1$} & \multirow{2}{*}{$6$} & \multirow{2}{*}{$n$} & \multirow{2}{*}{$3n-3$} & \multirow{2}{*}{$\frac{6n}{12n+1}$} & $12n+1$ prime, & \multirow{2}{*}{Theorem 12 \cite{MT}}\\
&&&&&$(-3)^{3n} \not\equiv 1 \pmod{12n+1}$. &\\
%\multirow{2}{*}{$20n+1$} & \multirow{2}{*}{$5$} & \multirow{2}{*}{$2n$} & \multirow{2}{*}{$5n-3$} & \multirow{2}{*}{$\frac{10n}{20n+1}$} & $20n+1$ prime, & \multirow{2}{*}{\cite{MT}}\\
%&&&&&$5^{5n} \not\equiv 1 \pmod{20n+1}$&\\
$8n+5$ & $2$ & $2n+1$ & $2n$ & $\frac{4n+2}{8n+5}$ & $8n+5$ prime & Theorem 3 \cite{MT}\\
$\frac{n^{2t+1}-1}{n-1}$ & $n+1$ & $\frac{n^{2t}-1}{n^2-1}$ & $\frac{n^{2t-1}-n}{n-1}$ & $\frac{n^{2t}-1}{n^{2t+1}-1}$ &
$\begin{cases}p=2, s \leq 5\\ p=3, s \leq 3\\ p=5, 8, 9, s = 2\end{cases}$.& \cite{FJV,FMT}\\
%$\frac{n^{2t+2}-1}{n-1}$ & $n^{t+1}$ & $\frac{n^t-1}{n-1}$ & $\frac{(n^t-n)(n^t-1)}{n-1}$ & $\frac{n^{t+1}(n^t-1)}{n^{2t+2}-1}$ & $n$ prime power, $t \geq 2$ & Corollary 4.2 \cite{FL}\\
\hline
\multirow{3}{*}{$16n+1$} & \multirow{3}{*}{$2n$} & \multirow{3}{*}{$4$} & \multirow{3}{*}{$3n$} & \multirow{3}{*}{$\frac{8n}{16n+1}$} &
$16n+1 = x^2+4y^2 = a^2+2b^2$ prime, & \multirow{3}{*}{Unpublished \cite{Mutoh}\rlap{\textsuperscript{b}}}\\
 & &  & & & $x \equiv a \equiv 1 \pmod{4}$, & \\
 & &  & & & $x+2a=3$. & \\
 \multirow{5}{*}{$20n+1$} & \multirow{5}{*}{$2n$} & \multirow{5}{*}{$5$} & \multirow{5}{*}{$4n$} & \multirow{5}{*}{$\frac{10n}{20n+1}$} &
$20n+1$ prime, & \multirow{5}{*}{Unpublished \cite{Mutoh}\rlap{\textsuperscript{c}}}\\
 & &  & & & $320n+16 = a^2+50b^2 +50c^2+125d^2$, & \\
 & &  & & & $a \equiv 1 \pmod{5}$, & \\
 & &  & & & $ad = c^2-4bc-b^2$, & \\
 & &  & & & $\begin{cases}a = -4 \text{\ if\ }2 \text{\ is a fifth power in\ } \mathbb{F}_{20n+1}\\a+20b-10c+25d=16 \text{\ otherwise}\end{cases}$. & \\
 \hline
 \hline
\multicolumn{7}{l}{\scriptsize\textsuperscript{a} These DSSs were also independently discovered in \cite{HW} as external difference families.}\vspace{-1mm}\\
\multicolumn{7}{l}{\scriptsize\textsuperscript{b} See Theorem 22 in \cite{T}.}\vspace{-1mm}\\
\multicolumn{7}{l}{\scriptsize\textsuperscript{c} See Theorem 23 in \cite{T}.}\vspace{2mm}
\end{tabular}
\end{table*}

\section*{Acknowledgment}
The authors thank Yukiyasu Mutoh for sharing his unpublished manuscript \cite{Mutoh},
and are grateful to the anonymous referees and Associate Editor Kyeongcheol Yang for careful reading and valuable comments.
This research was conducted while the first author was visiting the Department of Mathematical Sciences, Michigan Technological University.
He thanks the department for the hospitality.

% Generated by IEEEtranS.bst, version: 1.13 (2008/09/30)

\begin{IEEEbiographynophoto}{Yuichiro Fujiwara}
(M'10) received the B.S. and M.S. degrees in mathematics from Keio University, Japan,
and the Ph.D. degree in information science from Nagoya University, Japan.

He was a JSPS postdoctoral research fellow with the Graduate School of
System and Information Engineering, Tsukuba University, Japan, and a visiting
scholar with the Department of Mathematical Sciences, Michigan Technological University.
He is currently with the Division of Physics, Mathematics and Astronomy,
California Institute of Technology, Pasadena, where he works as a visiting postdoctoral research fellow.

Dr.\ Fujiwara's research interests include combinatorics and its interaction with computer science and quantum information science,
with particular emphasis on combinatorial design theory, algebraic coding theory, and quantum information theory.
\end{IEEEbiographynophoto}

\begin{IEEEbiographynophoto}{Vladimir D. Tonchev}
graduated with PhD in Mathematics from the University of Sofia,
Bulgaria, in 1980, and received the  Dr. of Mathematical Sciences
degree from the Bulgarian Academy of Sciences in 1987.
After spending a year as a research fellow at the Eindhoven University
of Technology, The Netherlands, (1987-88), and two years at the
universities of Munich, Heidelberg and Giessen in Germany as
an Alexander von Humboldt Research Fellow (1988-90), Dr. Tonchev joined
Michigan Technological University, where he is currently
a Professor of Mathematical Sciences.
Tonchev has published over 160 papers, four books,
three book chapters, and edited several volumes in the area of
error-correcting codes,
combinatorial  designs, and their applications.
Dr. Tonchev is a member of the editorial board of \textit{Designs, Codes and Cryptography},
\textit{Journal of Combinatorial Designs}, \textit{Applications and Applied Mathematics},
and \textit{Albanian Journal of Mathematics}, and a Founding Fellow of
the Institute of
Combinatorics and its Applications.
\end{IEEEbiographynophoto}


\begin{thebibliography}{10}
\providecommand{\url}[1]{#1}
\csname url@samestyle\endcsname
\providecommand{\newblock}{\relax}
\providecommand{\bibinfo}[2]{#2}
\providecommand{\BIBentrySTDinterwordspacing}{\spaceskip=0pt\relax}
\providecommand{\BIBentryALTinterwordstretchfactor}{4}
\providecommand{\BIBentryALTinterwordspacing}{\spaceskip=\fontdimen2\font plus
\BIBentryALTinterwordstretchfactor\fontdimen3\font minus
  \fontdimen4\font\relax}
\providecommand{\BIBforeignlanguage}[2]{{%
\expandafter\ifx\csname l@#1\endcsname\relax
\typeout{** WARNING: IEEEtranS.bst: No hyphenation pattern has been}%
\typeout{** loaded for the language `#1'. Using the pattern for}%
\typeout{** the default language instead.}%
\else
\language=\csname l@#1\endcsname
\fi
#2}}
\providecommand{\BIBdecl}{\relax}
\BIBdecl

\bibitem{CD}
Y.~Chang and C.~Ding, ``Constructions of external difference families and
  disjoint difference families,'' \emph{Des. Codes Cryptogr.}, vol.~40, pp.
  167--185, 2006.

\bibitem{CLY}
Y.~M. Chee, A.~C.~H. Ling, and J.~Yin, ``Optimal partitioned cyclic difference
  packings for frequency hopping and code synchronization,'' \emph{{IEEE}
  Trans. Inf. Theory}, vol.~56, pp. 5738--5746, 2010.

\bibitem{CC2}
W.~Chu and C.~J. Colbourn, ``Optimal frequency-hopping sequences via
  cyclotomy,'' \emph{{IEEE} Trans. Inf. Theory}, vol.~51, pp. 1139--1141, 2005.

\bibitem{CCD}
W.~Chu, C.~J. Colbourn, and P.~Dukes, ``On constant composition codes,''
  \emph{Discrete Appl. Math.}, vol. 154, pp. 912--929, 2006.

\bibitem{CHY}
J.-H. Chung, Y.~K. Han, and K.~Yang, ``New classes of optimal frequency-hopping
  sequences by interleaving techniques,'' \emph{{IEEE} Trans. Inf. Theory},
  vol.~55, pp. 5783--5791, 2009.

\bibitem{CY}
J.-H. Chung and K.~Yang, ``Optimal frequency-hopping sequences with new
  parameters,'' \emph{{IEEE} Trans. Inf. Theory}, vol.~56, pp. 1685--1693,
  2010.

\bibitem{CY2}
J.-H. Chung and K.~Yang, ``$k$-{Fold} cyclotomy and its application to frequency-hopping
  sequences,'' \emph{{IEEE} Trans. Inf. Theory}, vol.~57, pp. 2306--2317, 2011.

\bibitem{HandbookCD}
C.~J. Colbourn and J.~H. Dinitz, Eds., \emph{Handbook of {C}ombinatorial
  {D}esigns}, 2nd~ed.\hskip 1em plus 0.5em minus 0.4em\relax Boca Raton, FL:
  Chapman \& Hall/CRC, 2007.

\bibitem{Dickson}
L.~E. Dickson, ``Cyclotomy, higher congruences, and {Waring's} problem,''
  \emph{Amer. J. Math.}, vol.~57, pp. 391--424, 1935.

\bibitem{D}
C.~Ding, ``Optimal and perfect difference systems of sets,'' \emph{J. Combin.
  Theory Ser. A}, vol. 116, pp. 109--119, 2008.

\bibitem{Ding}
C.~Ding, ``Optimal constant composition codes from zero-difference balanced
  functions,'' \emph{{IEEE} Trans. Inf. Theory}, vol.~54, pp. 5766--5770, 2008.

\bibitem{DFFJM}
C.~Ding, R.~Fuji-Hara, Y.~Fujiwara, M.~Jimbo, and M.~Mishima, ``Sets of
  frequency hopping sequences: bounds and optimal constructions,'' \emph{{IEEE}
  Trans. Inf. Theory}, vol.~55, pp. 3297--3304, 2009.

\bibitem{DMY}
C.~Ding, M.~Miosio, and J.~Yuan, ``Algebraic constructions of optimal frequency
  hopping sequences,'' \emph{{IEEE} Trans. Inf. Theory}, vol.~53, pp.
  2606--2610, 2007.

\bibitem{DY}
C.~Ding and J.~Yin, ``Combinatorial constructions of optimal constant
  composition codes,'' \emph{{IEEE} Trans. Inf. Theory}, vol.~51, pp.
  3671--3673, 2005.

\bibitem{FL}
C.-L. Fan and J.-G. Lei, ``Constructions of difference systems of sets from
  finite projective geometry,'' \emph{{IEEE} Trans. Inf. Theory}, vol.~58,
  no.~1, pp. 130--138, 2012.

\bibitem{FJV}
R.~Fuji-Hara, M.~Jimbo, and S.~Vanstone, ``Some results on the line partitioning
  problem in PG$(2k,q)$,'' \emph{Util. Math.}, vol.~30, pp.
  235--241, 1986.

\bibitem{FMM}
R.~Fuji-Hara, Y.~Miao, and M.~Mishima, ``Optimal frequency hopping sequences: A
  combinatorial approach,'' \emph{{IEEE} Trans. Inf. Theory}, vol.~50, pp.
  1408--2420, 2004.

\bibitem{FMY}
R.~Fuji-Hara, K.~Momihara, and M.~Yamada, ``Perfect difference systems of sets
  and Jacobi sums,'' \emph{Discrete Math.}, vol. 309, pp. 3954--3961, 2009.

\bibitem{FMT}
R.~Fuji-Hara, A.~Munemasa, and V.~D. Tonchev, ``Hyperplane partitions and
  difference systems of sets,'' \emph{J. Combin. Theory Ser. A}, vol. 113, pp.
  1699--1718, 2006.

\bibitem{FF}
Y.~Fujiwara and R.~Fuji-Hara, ``Frequency hopping sequences with optimal auto-
  and cross-correlation properties and related codes,'' in \emph{Proc. Tenth
  Int. Workshop Algebraic and Combin. Coding Theory}, vol.~10, 2006, pp.
  93--96.

\bibitem{GFM}
G.~Ge, R.~Fuji-Hara, and Y.~Miao, ``Further combinatorial constructions for
  optimal frequency hopping sequences,'' \emph{J. Combin. Theory Ser. A}, vol.
  113, pp. 1699--1718, 2006.

\bibitem{GMW}
G.~Ge, Y.~Miao, and L.~Wang, ``Combinatorial constructions for optimal
  splitting authentication codes,'' \emph{{SIAM} J. Discrete Math.}, vol.~18,
  pp. 663--678, 2005.

\bibitem{HY}
Y.~K. Han and K.~Yang, ``On the {Sidel'nikov} sequences as frequency-hopping
  sequences,'' \emph{{IEEE} Trans. Inf. Theory}, vol.~55, pp. 4279--4285, 2009.

\bibitem{HW}
B.~Huang and D.~Wu, ``Cyclotomic constructions of external difference families and disjoint difference families,''
  \emph{J. Combin. Des.}, vol.~17, pp. 333--341, 2009.

\bibitem{KL}
J.~J. Komo and S.~C. Liu, ``Maximal length sequences for frequency hopping,''
  \emph{{IEEE} J. Sel. Areas Commun.}, vol.~5, pp. 819--822, 1990.

\bibitem{Kumar}
P.~V. Kumar, ``Frequency-hopping code sequence designs having large linear
  span,'' \emph{{IEEE} Trans. Inf. Theory}, vol.~34, pp. 146--151, 1988.

\bibitem{LF}
J.~Lei and C.~Fan, ``Optimal difference systems of sets and partition-type
  cyclic difference packings,'' \emph{Des. Codes Cryptogr.}, vol.~58, pp.
  135--153, 2011.

\bibitem{LG}
A.~Lempel and H.~Greenberger, ``Families of sequences with optimal hamming
  correlation properties,'' \emph{{IEEE} Trans. Inf. Theory}, vol.~20, pp.
  90--94, 1974.

\bibitem{L}
V.~I. Levenshtein, ``One method of constructing quasi codes providing
  synchronization in the presence of errors,'' \emph{Problems Inform.
  Transmission}, vol.~7, no.~3, pp. 215--222, 1971.

\bibitem{L2}
V.~I. Levenshtein, ``Combinatorial problems motivated by comma-free codes,'' \emph{J.
  Combin. Des.}, vol.~12, pp. 184--196, 2004.

\bibitem{LH}
J.~Luo and T.~Helleseth, ``Constant composition codes as subcodes of cyclic
  codes,'' \emph{{IEEE} Trans. Inf. Theory}, vol.~57, pp. 7482--7488, 2011.

\bibitem{LFVC}
Y.~Luo, F.~W. Fu, A.~J. {Han Vinck}, and W.~Chen, ``On constant composition
  codes over ${\textit{\textbf{z}}}_p$,'' \emph{{IEEE} Trans. Inf. Theory},
  vol.~49, pp. 3010--3016, 2003.

\bibitem{PCMutoh}
Y.~Mutoh, \emph{private communication}.

\bibitem{Mutoh}
Y.~Mutoh, ``Difference systems of sets and cyclotomy {II},'' \emph{unpublished
  manuscript}.

\bibitem{MT}
Y.~Mutoh and V.~D. Tonchev, ``Difference systems of sets and cyclotomy,''
  \emph{Discrete Math.}, vol. 308, pp. 2959--2969, 2008.

\bibitem{OKSS}
W.~Ogata, K.~Kurosawa, D.~R. Stinson, and H.~Saido, ``New combinatorial designs
  and their applications to authentication codes and secret sharing schemes,''
  \emph{Discrete Math.}, vol. 279, pp. 383--405, 2004.

\bibitem{T2}
V.~D. Tonchev, ``Difference systems of sets and code synchronization,''
  \emph{Rendiconti del Seminario Matematico di Messina Series {II}}, vol.~9,
  pp. 217--226, 2003.

\bibitem{T}
V.~D. Tonchev, ``Partitions of difference sets and code synchronization,''
  \emph{Finite Fields Appl.}, vol.~11, pp. 601--621, 2005.
  
\bibitem{web}
V.~D. Tonchev, ``Tables of DSS for $q = 2$, $3$, $4$.''
  [Online]. Available:
  \url{http://www.math.mtu.edu/~tonchev/DSS.htm}

\bibitem{US}
P.~Udaya and M.~N. Siddiqi, ``Optimal large linear complexity frequency hopping
  patterns derived from polynomial residue class rings,'' \emph{{IEEE} Trans.
  Inf. Theory}, vol.~44, pp. 1492--1503, 1998.

\bibitem{W}
H.~Wang, ``A new bound for difference systems of sets,'' \emph{J. Combin. Math.
  Combin. Comput.}, vol.~58, pp. 161--168, 2006.

\bibitem{WW}
X.~Wang and J.~Wang, ``Optimal difference systems of sets and difference
  sets,'' \emph{Aequat. Math.}, vol.~82, pp. 155--164, 2011.

\bibitem{WW2}
X.~Wang and J.~Wang, ``Partitioned difference families and almost difference sets,''
  \emph{J. Statist. Plann. Infer.}, vol. 141, pp. 1899--1909, 2011.

\bibitem{YTPP}
Y.~Yang, X.~Tang, U.~Parampalli, and D.~Peng, ``New bound on frequency hopping
  sequence sets and its optimal constructions,'' \emph{{IEEE} Trans. Inf.
  Theory}, vol.~57, pp. 7605--7613, 2011.

\bibitem{YST}
J.~Yin, X.~Shan, and Z.~Tian, ``Constructions of partitioned difference
  families,'' \emph{European J. Combin.}, vol.~29, pp. 1507--1519, 2008.

\bibitem{ZCTY}
X.~Zeng, H.~Cai, X.~Tang, and Y.~Yang, ``A class of optimal frequency hopping
  sequences with new parameters,'' \emph{{IEEE} Trans. Inf. Theory}, vol.~58,
  pp. 4899--4907, 2012.

\bibitem{ZTPP}
Z.~Zhou, X.~Tang, D.~Peng, and U.~Parampalli, ``New constructions for optimal
  sets of frequency-hopping sequences,'' \emph{{IEEE} Trans. Inf. Theory},
  vol.~57, pp. 3831--3840, 2011.

\bibitem{ZTWY}
Z.~Zhou, X.~Tang, D.~Wu, and Y.~Yang, ``Some new classes of zero-difference
  balanced functions,'' \emph{{IEEE} Trans. Inf. Theory}, vol.~58, pp.
  139--145, 2012.

\bibitem{ZG}
M.~Zhu and G.~Ge, ``Quaternary constant-composition codes with weight $4$ and
  distances $5$ or $6$,'' \emph{{IEEE} Trans. Inf. Theory}, vol.~58, pp.
  6012--6022, 2012.

\end{thebibliography}
\end{document}